\documentclass{ifacconf}

\usepackage{graphicx}      

\makeatletter
\let\old@ssect\@ssect 
\makeatother

\usepackage{natbib}
\usepackage{hyperref}

\makeatletter
\def\@ssect#1#2#3#4#5#6{%
	\NR@gettitle{#6}
	\old@ssect{#1}{#2}{#3}{#4}{#5}{#6}
}
\makeatother

\newcommand{\cost}{\Psi}

\usepackage[T1]{fontenc} 

\usepackage[utf8]{inputenc} 

\usepackage{array,booktabs}
\usepackage{subfig}
\usepackage{makecell}
\usepackage{float}
\usepackage{pdfpages}
\usepackage{subfig} 

\usepackage{amsmath,amssymb} 
\usepackage{mathtools}
\usepackage{amsfonts}
\usepackage{bbm}
\usepackage{multirow}
\usepackage{lastpage}
\usepackage{scrtime}
\usepackage{mathrsfs,dsfont} 
\usepackage{eurosym}
\usepackage{amscd}
\usepackage{caption}
\usepackage{varioref} 
\usepackage{tikz}
\usetikzlibrary {positioning}
\usepackage{pgfplots}
\pgfplotsset{ 
	compat=newest, 
	legend style =
	{font=\small \sffamily},
	label style = {font=\small\sffamily},
	every tick label/.append style={font=\small}}
\newcommand{\mc}{\mathcal}
\usepackage[most,skins,theorems]{tcolorbox}
\tcbset{highlight math style={enhanced,
		colframe=red,colback=white,arc=0pt,boxrule=1pt}}
\usetikzlibrary{arrows}

\newcommand{\abs}[1]{|#1|}
\def\1{\mathds{1}}

\def\R{\mathbb{R}}

\newcommand{\overbar}[1]{\mkern 1.5mu\overline{\mkern-1.5mu#1\mkern-1.5mu}\mkern 1.5mu}

\usetikzlibrary[topaths]
\newcount\mycount
\DeclareMathOperator*{\argmax}{arg\,max}

\newtheorem{corollary}{Corollary}
\usepackage[shortlabels]{enumitem}

\usepackage[framemethod=tikz]{mdframed}

\usepackage{xparse}

\def\exampletext{Example} 

\NewDocumentEnvironment{testexample}{ O{} }
{
	\colorlet{colexam}{red!55!black} 
	\newtcolorbox[use counter=testexample]{testexamplebox}{%
		empty,
		title={\exampletext: #1},
		attach boxed title to top left,
		minipage boxed title,
		boxed title style={empty,size=minimal,toprule=0pt,top=4pt,left=3mm,overlay={}},
		coltitle=colexam,fonttitle=\bfseries,
		before=\par\medskip\noindent,parbox=false,boxsep=0pt,left=3mm,right=0mm,top=2pt,breakable,pad at break=0mm,
		before upper=\csname @totalleftmargin\endcsname0pt, 
		overlay unbroken={\draw[colexam,line width=.5pt] ([xshift=-0pt]title.north west) -- ([xshift=-0pt]frame.south west); },
		overlay first={\draw[colexam,line width=.5pt] ([xshift=-0pt]title.north west) -- ([xshift=-0pt]frame.south west); },
		overlay middle={\draw[colexam,line width=.5pt] ([xshift=-0pt]frame.north west) -- ([xshift=-0pt]frame.south west); },
		overlay last={\draw[colexam,line width=.5pt] ([xshift=-0pt]frame.north west) -- ([xshift=-0pt]frame.south west); },%
	}
	\begin{testexamplebox}}
	{\end{testexamplebox}\endlist}

\usepackage[framemethod=tikz]{mdframed}

\definecolor{mycolor}{rgb}{0.122, 0.435, 0.698}

\newmdenv[innerlinewidth=0.5pt, roundcorner=4pt,linecolor=mycolor,innerleftmargin=6pt,
innerrightmargin=6pt,innertopmargin=6pt,innerbottommargin=6pt]{mybox}
\AtBeginDocument{
	\label{CorrectFirstPageLabel}
	
}

\begin{document}
\begin{frontmatter}

\title{Equilibria in Network Constrained Energy Markets\thanksref{footnoteinfo}} 

\thanks[footnoteinfo]{This work was supported by Ministero dell'Istruzione, dell'Universita e della Ricerca [Grant 
	E11G18000350001 and Research Project PRIN 2017 “Advanced Network Control of Future Smart
	Grids”] and the Compagnia di San Paolo.}

\author[First]{Giacomo Como} 
\author[First]{Fabio Fagnani} 
\author[First]{Leonardo Massai}

\address[First]{Department of Mathematical Sciences ``G.L.~Lagrange'', Politecnico di Torino, Corso Duca degli Abruzzi 24, 10129 Torino, Italy\\ 
	(e-mail: \{leonardo.massai,giacomo.como,fabio.fagnani\}@polito.it).}

\begin{abstract}                
We study an energy market composed of producers who compete to supply energy to different markets and want to maximize their profits. The energy market is modeled by a graph representing a constrained power network where nodes represent the markets and links are the physical lines with a finite capacity connecting them. Producers play a networked Cournot game on such a network together with a centralized authority, called market maker, that facilitates the trade between geographically separate markets via the constrained power network and aims to maximize a certain welfare function. We first prove a general result that links the optimal action of the market maker with the capacity constraint enforced on the power network. Under mild assumptions, we study the existence and uniqueness of Nash equilibria and exploit our general result to prove a connection between capacity bottlenecks in the power network and the emergence of price differences between different markets that are separated by saturated lines, a phenomenon that is often observed in real power networks.
\end{abstract}

\begin{keyword}
 Game theory, Energy systems, Networked systems, Game theory for natural resources, Power systems.
\end{keyword}

\end{frontmatter}

\section{Introduction}
The studying of network effects in modern marketplaces has attracted a considerable amount of attention in recent years.  In particular, a growing body of literature has pointed out how classical models of competition that often feature several producers operating in a single, isolated market, fail to capture the growing interconnectedness that characterizes power systems, transportation and infrastructure networks and so on. 
These complex interconnections among different agents turn out to be
crucial to properly modeling and understanding emergent features of modern marketplaces. Consequently, several works in literature are devoted to studying networked models of competition. In \cite{a14, Bimpikis2014} authors extend the classical model of Cournot competition by considering multiple firms operating in different markets. In this setting, a sort of bipartite graph arises, coupling producers and markets via the non separability of each producer's cost function in the markets it participates in.

\medskip

Other works have focused on specific applications, for instance electricity power models  where the physical network connecting different markets is fundamental.  In \cite{Barquin2005} and \cite{Barquin8} a constrained power network connecting different markets and producers following a Cournot competition scheme is considered and the authors develop an iterative  algorithm for finding the Nash equilibrium, which considers how the production at a certain node affects the whole network, and consequently explains the opportunities for the producers of exercising market power. In \cite{Neuhoff2005} a numerical estimation of how sensitive Nash equilibria are in a networked Cournot competition in a transmission-constrained electricity market is performed, highlighting that Cournot equilibria are indeed highly sensitive to assumptions about market design. A  two-settlement electricity markets with the forward market and the spot market is introduced in \cite{Yao2008}, which accounts for flow congestion, demand uncertainty, system contingencies, and market power. The model assumes linear demand functions, quadratic generation cost functions, and a lossless DC power network.

Our paper fits into this growing literature on networked Cournot competition and, in particular, takes cue from the work of Cai \etal  \cite{Cai2019} where networked Cournot competition among multiple energy producers is studied together with the presence of an additional player called market maker, a centralized authority that moves supply between geographically separate markets via the constrained power network to achieve a desirable state of the system. Their focus is on understanding the consequences of the design of the market maker utility function and providing tools for optimal design.

\medskip

We should mention that the Cournot competition tailored to energy markets usually contrasts with other popular Cournot game schemes where producers decide their production quantities (a vector) over a whole set of markets. In this
latter scheme, the producers readily consider both
production and distribution of a certain good in their
decision-making process, and thus no market maker is introduced into the game. The aforementioned scheme is considered for example in \cite{2020p, 2021g, 9130079}. However, this kind of competition without intermediaries nor market-to-market energy exchange is not realistic when modeling an energy market. The complexity and the broad impact of energy marketplaces on the whole environmental and economic policy of a government typically leads to (and often necessitates) the emergence of intermediaries. In these markets, a centralized authority typically solves a dispatch problem by utilizing the offers/bids
from the generators/retailers and aims to maximize some metric of social welfare subject to the operational constraints of the grid. This is exactly the model that we want to capture by  considering a market maker that plays a specific role with respect to transport and trade of energy between markets. By doing so, the market maker is also a key figure in
matching the demand and supply of power and, as an independent regulated entity, it further designs rules, via the choice of its utility function, to limit the possible exercise of market power by the producers. All these aspects cannot be modeled if producers are in charge of the total quantity supplied as well as distribution.

\medskip

In this paper, we model the energy market by means of a graph that represents a constrained power network  where nodes represent the markets and links are the physical lines with a finite capacity connecting them. Producers play a networked Cournot game on such a network together with a market maker that aims to procure supply from one market and transport it to a different market in order to maximize a certain welfare function.  In contrast to \cite{Cai2019}, we weaken some assumptions on the price and cost functions and we do not focus on the design of the market maker's utility function but rather on the studying the Nash equilibria of the game and on highlighting the impact of the capacity constraints on such equilibria. More specifically, our main result holds under extremely mild assumptions on the market maker's utility function and it establishes a fundamental connection between the optimal action of the market maker and the capacity constraint. We proceed by increasingly adding more structure  on the market maker and producers' utility functions. To begin with, we prove existence of Nash equilibria under standard concavity hypotheses; moreover, when the market maker's utility function is equal to the well studied Marshallian welfare (see \cite{Johari2005}), we prove that the considered game is potential and admits a unique Nash equilibrium that can be efficiently found by solving a concave optimization problem. 
In this more particular setting, our main result establishes that, at equilibrium, if there is a mismatch between prices at different markets this implies the existence of a saturated cut in the network dividing those markets, i.e, there exists a set of links connecting markets with different prices that are at full capacity at equilibrium. This result formally proves a connection between price differences and capacity constraints, a phenomenon that is often documented in real-world power networks \cite{who}, \cite{who2}.

%
%

\medskip

The rest of the paper is organized as follows. The reminder
of this section is devoted to the introduction of some
notational conventions used throughout the paper.
In \autoref{sec2} we present the model of networked Cournot competition on power networks that is the object of our study. In \autoref{sec3} we present our findings, starting with our main contribution that, under very mild assumptions on the market maker's utility function, establishes  a key connection between optimal actions of the market maker and saturated cuts in the power network. Afterward, we add a few standard hypotheses on the producers and market maker's utility functions and this allows us to prove results
concerning the existence and uniqueness of the Nash equilibria. Moreover, in this particular setting we prove a Corollary of our main result that links the emergence of price differences with capacity bottlenecks in the power network at equilibrium. This Section is complemented with an Example that shows our results on a simplified Italian power network model. Finally, in \autoref{sec4} we draw some conclusions and discuss current and future research.

\medskip

Throughout the paper we shall denote vectors with lower case, matrices with upper case,
and sets with calligraphic letters. We indicate with $\1$ the all-1 vector and with $I$ the identity matrix, regardless of their dimension. Moreover, given a path $\gamma$ in a directed graph with $l$ links, we will denote with $\1_{\gamma} \in \{0,1\}^l$ the vector  whose generic component is equal to 1 if and only if the link associated to that component belongs to the path $\gamma$. Finally, we denote with $\delta_j$ the $j$-th component of the canonical basis of $\R^n$.



\section{The Model} \label{sec2}
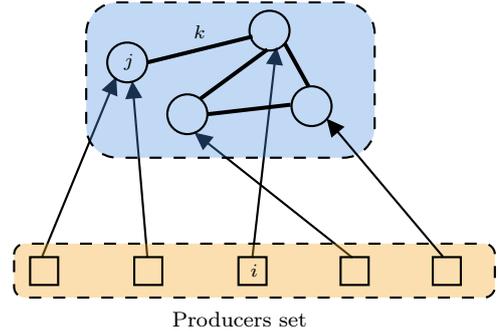
\begin{figure}[t]
	\centering
	\begin{tikzpicture}[thick, x=0.75pt,y=0.75pt,yscale=-1,xscale=1]
		
		\draw    (241.25,121.32) -- (248.5,206.74) ;
		\draw [shift={(241,118.33)}, rotate = 85.15] [fill={rgb, 255:red, 0; green, 0; blue, 0 }  ][line width=0.08]  [draw opacity=0] (8.93,-4.29) -- (0,0) -- (8.93,4.29) -- cycle    ;
		\draw    (231.86,119.11) -- (196,206.74) ;
		\draw [shift={(233,116.33)}, rotate = 112.26] [fill={rgb, 255:red, 0; green, 0; blue, 0 }  ][line width=0.08]  [draw opacity=0] (8.93,-4.29) -- (0,0) -- (8.93,4.29) -- cycle    ;
		\draw    (312.66,104.31) -- (301,206.74) ;
		\draw [shift={(313,101.33)}, rotate = 96.49] [fill={rgb, 255:red, 0; green, 0; blue, 0 }  ][line width=0.08]  [draw opacity=0] (8.93,-4.29) -- (0,0) -- (8.93,4.29) -- cycle    ;
		\draw    (274.35,146.2) -- (351,207.33) ;
		\draw [shift={(272,144.33)}, rotate = 38.57] [fill={rgb, 255:red, 0; green, 0; blue, 0 }  ][line width=0.08]  [draw opacity=0] (8.93,-4.29) -- (0,0) -- (8.93,4.29) -- cycle    ;
		\draw  [fill={rgb, 255:red, 74; green, 144; blue, 226 }  ,fill opacity=0.34 ][dash pattern={on 4.5pt off 4.5pt}] (218,94.49) .. controls (218,85.86) and (225,78.86) .. (233.63,78.86) -- (346.37,78.86) .. controls (355,78.86) and (362,85.86) .. (362,94.49) -- (362,141.37) .. controls (362,150) and (355,157) .. (346.37,157) -- (233.63,157) .. controls (225,157) and (218,150) .. (218,141.37) -- cycle ;
		\draw  [fill={rgb, 255:red, 245; green, 166; blue, 35 }  ,fill opacity=0.36 ][dash pattern={on 4.5pt off 4.5pt}] (182,205.65) .. controls (182,202.66) and (184.43,200.23) .. (187.42,200.23) -- (416.58,200.23) .. controls (419.57,200.23) and (422,202.66) .. (422,205.65) -- (422,221.91) .. controls (422,224.91) and (419.57,227.33) .. (416.58,227.33) -- (187.42,227.33) .. controls (184.43,227.33) and (182,224.91) .. (182,221.91) -- cycle ;
		\draw    (396,206.67) -- (339.92,139.63) ;
		\draw [shift={(338,137.33)}, rotate = 50.09] [fill={rgb, 255:red, 0; green, 0; blue, 0 }  ][line width=0.08]  [draw opacity=0] (8.93,-4.29) -- (0,0) -- (8.93,4.29) -- cycle    ;
		\draw [line width=1.5]    (301,96.05) -- (248.5,109.07) ;
		\draw [line width=1.5]    (317,99.33) -- (329,121.1) ;
		\draw [line width=1.5]    (308.5,102.56) -- (275,127.33) ;
		\draw [line width=1.5]    (320,130.33) -- (278.5,135.12) ;
		\draw   (228.5,109.07) .. controls (228.5,103.55) and (232.98,99.07) .. (238.5,99.07) .. controls (244.02,99.07) and (248.5,103.55) .. (248.5,109.07) .. controls (248.5,114.6) and (244.02,119.07) .. (238.5,119.07) .. controls (232.98,119.07) and (228.5,114.6) .. (228.5,109.07) -- cycle ;
		\draw   (258.5,135.12) .. controls (258.5,129.6) and (262.98,125.12) .. (268.5,125.12) .. controls (274.02,125.12) and (278.5,129.6) .. (278.5,135.12) .. controls (278.5,140.64) and (274.02,145.12) .. (268.5,145.12) .. controls (262.98,145.12) and (258.5,140.64) .. (258.5,135.12) -- cycle ;
		\draw   (299.5,93.07) .. controls (299.5,87.55) and (303.98,83.07) .. (309.5,83.07) .. controls (315.02,83.07) and (319.5,87.55) .. (319.5,93.07) .. controls (319.5,98.6) and (315.02,103.07) .. (309.5,103.07) .. controls (303.98,103.07) and (299.5,98.6) .. (299.5,93.07) -- cycle ;
		\draw   (320.5,131.07) .. controls (320.5,125.55) and (324.98,121.07) .. (330.5,121.07) .. controls (336.02,121.07) and (340.5,125.55) .. (340.5,131.07) .. controls (340.5,136.6) and (336.02,141.07) .. (330.5,141.07) .. controls (324.98,141.07) and (320.5,136.6) .. (320.5,131.07) -- cycle ;
		\draw   (190,207) -- (204,207) -- (204,221) -- (190,221) -- cycle ;
		\draw   (242,207) -- (256,207) -- (256,221) -- (242,221) -- cycle ;
		\draw   (294,207) -- (308,207) -- (308,221) -- (294,221) -- cycle ;
		\draw   (345,207) -- (359,207) -- (359,221) -- (345,221) -- cycle ;
		\draw   (391,207) -- (405,207) -- (405,221) -- (391,221) -- cycle ;
		
		\draw (172.88,61.04) node [anchor=north west][inner sep=0.75pt]  [font=\footnotesize] [align=left] {Markets set and constrained power network \ };
		\draw (259.38,233.27) node [anchor=north west][inner sep=0.75pt]  [font=\footnotesize] [align=left] {Producers set };
		\draw (235.38,103.17) node [anchor=north west][inner sep=0.75pt]  [font=\scriptsize]  {$j$};
		\draw (270.13,89.15) node [anchor=north west][inner sep=0.75pt]  [font=\scriptsize]  {$k$};
		\draw (298.38,209.17) node [anchor=north west][inner sep=0.75pt]  [font=\scriptsize]  {$i$};

	\end{tikzpicture}

	\caption{Representation of an energy market with a constrained power network linking different markets.}
	\label{fig1}
\end{figure}

We consider $n$ competing producers that choose a production quantity $q_i$ with production cost functions $\cost_{i}\left(q_{i}\right), 1 \leq i \leq n$ and $m$ markets with total consumption $d_j$ and price function $P_{j}\left(d_{j}\right), 1 \leq j \leq m$. The markets are connected by $l$ links with finite capacity $c_{k}, 1 \leq k \leq l$. We assume that the market network forms a connected graph, in other words, there are no isolated markets. In Fig. \ref{fig1} we represent the constrained power network between the markets and the producers that are linked to them as a sort of bipartite graph. We collect the production quantities in the vector $q \in \R_+^n$, the capacities in $c \in \R_+^l$ while the vector $f \in \R^l$ is the flow of production quantity that the market maker moves around the network.

The network model can be described by means of two matrices: $B \in\{0, \pm 1\}^{m \times l}$ is the node-link incidence matrix (with arbitrary orientation) and $H \in\{0,1\}^{n \times m}$ is the producer-market incidence matrix.  With this in mind, the total consumption vector $d \in \R^m$, i.e., the vector collecting the total quantity of energy consumed in each market, can be simply written as $d=Bf+H^{\top}q$ where $r=Bf$ is the quantity moved in$\backslash$out of the markets by the market maker and $H^{\top}q$ is the total quantity made by all producers in each market.

The competition is modeled as a game with $ n+1 $ players (the $ n $ producers plus the market maker) where:  every producer $i=1,\ldots, n$ chooses to produce a quantity of energy $ q_i \ge 0 $ aiming at maximizing its utility 

\begin{equation}\label{eq1}
	u_{i}\left(q_{i}, q_{-i}, f\right)=q_{i} \sum_{j=1}^m H_{i j} P_{j}\left(\left(B f+H^{\top} q\right)_{j}\right)-\cost_{i}\left(q_{i}\right) \: ,
\end{equation}

the market maker chooses a flow vector $f \in \mathbb{R}^{l}$ satisfying the capacity constraints $|f| \leq c$ aiming at maximizing its utility
\begin{equation}\label{eq2}
	w: \R^l \times \R_+^n \mapsto \R \: , \: \quad w \in \mc C^1\: .
\end{equation}

It is useful to also define the utility of the market maker in the case when it depends on $f$ only through the term $r=Bf$, an assumption that we will use later on. In such a case we write the utility as a function $v$ such that 
\begin{equation}\label{eq3}
	v: \R^m \times \R_+^n \mapsto \R \: , \qquad v(r,q)=w(f,q) \: \forall \: f,q \: .
\end{equation}

Notice that the utility \eqref{eq1} of every producer represents its net profit (i.e., the difference between the total revenue $q_{i} \sum_{j=1}^m H_{i j} P_{j}\left(\left(B f+H^{\top} q\right)_{j}\right)$ and the production cost $\cost_{i}\left(q_{i}\right)$). 
On the other hand, notice that we are not specifying any particular  utility for the market maker and at this point we only require it to be a differentiable function of $f$ and $q$.

\smallskip


Throughout, we shall refer to the model described above as a networked Cournot game with market maker.  A (Nash) equilibrium for this game is a tuple $ (q^*,f^*) $  such that 
$$
\begin{aligned}
	u_{i}\left(q_{i}^*, q_{-i}^*, f^*\right) & \geq u_{i}\left(q_{i}, q_{-i}^*, f^*\right), \text { for all } q_{i} \in \mathbb{R}_{+} \\
	w(f^*, q^*) & \geq w(f, q^*), \text { for all } {f\in \R^l} \text{ such that } \abs{f}\le c \: .
\end{aligned}
$$
we shall also write  $r^*=Bf^*$ when we make use of the function $v$ defined in \eqref{eq3}.

\section{Main results} \label{sec3}
In this section we present the main results of this paper.
We start by stating our most general result that, as mentioned before, establishes a key connection between the optimal action made by the market maker and the capacity constraint.

\medskip

\begin{figure}[t]
	\centering
	\includegraphics[scale=0.6]{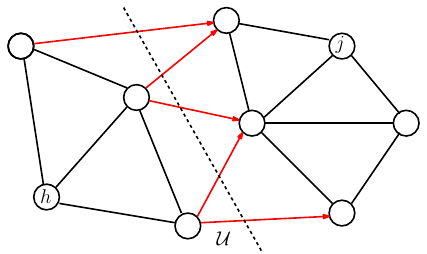}
	\caption{A cut in a network.}
	\label{fig2}
\end{figure}

\begin{thm}\label{th2}
	Assume that $w(f,q)$ depends on $f$ only through the term $r=Bf$ so that we can make use of the function $v$ defined in \eqref{eq3}, and $\forall q \in \R_+^n$
	\begin{align}& f^*  =\argmax_{0 \le f \le c} v(r, q) =\argmax_{0 \le f \le c} w(f, q)  \label{t1} \\[2ex]
		&	\exists \:  i,j \text{ such that }  \frac{\partial }{\partial r_i}v(r^*,q) <  \frac{\partial }{\partial r_j}v(r^*,q)	\label{t2}	
	\end{align}
	
	then,
	\begin{enumerate}
		\item 	it exists a saturated $i-j$ cut, i.e., there exists $\mathcal{U} \subseteq\{1, \ldots, m\}$ such that $$h \in \mathcal{U}, j \notin \mathcal{U} \quad f_{k}^{*}=\left\{\begin{array}{lll}c_{k} & \text { if } & k \text { from } \mathcal{U} \text { to } \mathcal{U}^{c} \\ -c_{k} & \text { if } & k \text { from } \mathcal{U}^{c} \text { to } \mathcal{U}\end{array}\right\} \: .$$ Here $\mathcal{U}^{c}$ denotes the complement of the set $\mc U$ (see Fig. \ref{fig2}).
		
		\item There is no flow on any $j-i$ path.
	\end{enumerate}

\end{thm}

\medskip

\begin{pf}
	
	\begin{enumerate}
		\item 	
		Let $f^*$ be such that \eqref{t1} holds. Assume that no $i-j$ cut is saturated, then, it exists a path $\gamma$ from $i$ to $j$ and a value $\overbar{\varepsilon}>0$ such that $f^*+\varepsilon \1_{\gamma} \le c \quad \forall \: \varepsilon \text{ such that } 0 \le \varepsilon \le \overbar{\varepsilon}$. 
		
		Now, taking the derivative with respect to $\varepsilon$ yields:
		
		\begin{align}
			& \left. \frac{\partial}{\partial \varepsilon} v(B\left(f^*+\varepsilon\1_{\gamma}\right), q) \right|_{\varepsilon=0}\\
			& = 	\left. \frac{\partial}{\partial \varepsilon} v\left(r^*+\varepsilon(\delta_j-\delta_i), q\right)\right|_{\varepsilon=0}\\
			& = \left(\delta_j-\delta_i\right)^{\top} \left. \nabla_r v\left(r^*+\varepsilon(\delta_j-\delta_i), q\right) \right|_{\varepsilon=0} \\
			&= \frac{\partial}{\partial r_j} v\left(r^*, q\right)-\frac{\partial}{\partial r_i} v\left(r^*, q\right)>0 \label{a1} \: .
		\end{align}
		
		Where we have used the chain rule and the fact that $B\1_{\gamma}=\delta_j-\delta_i$.

		The latter inequality proves that we can find a $\varepsilon^*>0$ such that $w(f^*,q) < w(f^*+\varepsilon^*\1_{\gamma},q)$ while still satisfying the capacity constraint $0\le f^*+\varepsilon^*\1_{\gamma} \le c$, which contradicts hypothesis \eqref{t1}, hence, a saturated $i-j$ cut must exist.  
		\item Assume that it exists a $j-i$ path $\gamma$ with positive flow on it. This means that we can always take a value $\varepsilon>0$ such that $f^*-\varepsilon\1_{\gamma} \ge 0$. Then, following the same argument used for proving point 1) we can write:
		
		\begin{align}
			& \left. \frac{\partial}{\partial \varepsilon} v(B\left(f^*-\varepsilon\1_{\gamma}\right), q) \right|_{\varepsilon=0}\\
			& = 	\left. \frac{\partial}{\partial \varepsilon} v\left(r^*-\varepsilon(\delta_i-\delta_j), q\right)\right|_{\varepsilon=0}\\
			& = \left(\delta_j-\delta_i\right)^{\top} \left. \nabla_r v\left(r^*+\varepsilon(\delta_j-\delta_i), q\right) \right|_{\varepsilon=0} \\
			&= \frac{\partial}{\partial r_j} v\left(r^*, q\right)-\frac{\partial}{\partial r_i} v\left(r^*, q\right)>0 \: .
		\end{align}
		This proves that we can find a $\varepsilon^*>0$ such that $w(f^*,q) < w(f^*-\varepsilon^*\1_{\gamma},q)$ , which contradicts again hypothesis \eqref{t1}, hence, no path $j-i$ with positive flow can exist.  
		
	\end{enumerate}
\end{pf}

We can notice that Theorem \ref{th2} is essentially a technical result that deals with the (not necessarily unique) maximizers $f^*$ of $w$ and saturated cuts in the network described by the matrix $B$. This result requires little to no assumption on the function $w$ and can indeed be completely decoupled from the game theoretical aspect of the model (notice that the utility of producers plays no role). It states that whenever the market maker "plays" the best move $f^*$ by maximizing its utility and that creates a mismatch in values between derivatives of the utility with respect to the total quantity $r$ injected into some nodes, then there must be a cut dividing those nodes consisting of links that are at maximum capacity under the optimal flow $f^*$.
This is particularly relevant as it poses specific restrictions on the flow chosen by the market maker given a certain capacity constraint and could be relevant for the optimal design of the utility function $w$ by the market maker itself.  Moreover, condition \eqref{t2} has a very natural interpretation in terms of a price mismatch when we chose a  commonly used form for $w$ as we will show in the following.

\medskip

We are now ready to add a few standard assumptions to the model. This allows us to study existence and uniqueness of Nash equilibria as well as to specialize Theorem \ref{th2} for a particular welfare function $w$.

\smallskip

More in details, the model is studied under the following assumptions on the cost/price functions and producers to markets relationships:
\begin{enumerate}[(a)]
	\item For all $1\le i \le n$ and $1 \le j \le m$, $\cost_{i}$ and $P_{j}$ are $\mathcal{C}^{2}$ with $\cost_{i}^{\prime} \geq 0, \cost_{i}^{\prime \prime} \geq 0, P_{j}^{\prime}<0, P_{j}^{\prime \prime}\le 0$ and it exists $\bar{q} \in (0,+\infty)$ such that $P_j(\bar{q})=0$.
	\item Each producer sells on single market: $\sum_{j=1}^m H_{i j}=1$.
	\item The function $w$ is given by
	\begin{equation}\label{eqmm}
		w(f, q)=\sum_{j=1}^m \int_{0}^{\left(B f+H^{\top} q\right)_{j}} P_{j}(s) \: \mathrm{d} s-\sum_{i=1}^n \cost_{i}\left(q_{i}\right) \: .
	\end{equation}
\end{enumerate}
Assumption (a) collects some standard conditions on regularity and concavity of functions $\cost$ and $P$.

A few comments on assumption (b) are in order. Although having producers selling on a single market may appear quite restrictive, this is actually what happens in most energy marketplaces. In Italy for instance, there are essentially three big markets for electricity (north, center and south) and producers make offers/bids in that specific market. The actual dispatch of energy between markets is operated by a centralized authority, in this case the market maker. From a technical perspective, dropping assumption (b) makes it unclear whether the resulting game remains potential (see Theorem \ref{th1}).

Finally, assumption (c) gives us a specific form for the utility $w$ to work with. This corresponds to the so-called Marshallian welfare that is widely used in this framework (see \cite{Johari2005}) and can be interpreted as the difference between the aggregate consumer surplus and the total production cost. Notice that this specific form for the market maker's utility only depends on $f$ through the term $r=Bf$ and can be written as a function $v(r,q)$ as defined in \eqref{eq3}.

\smallskip

Now that we have stated the main hypotheses, we are ready to present the following result that deals with existence and uniqueness of equilibria.

\medskip

\begin{thm}
	\label{th1}

	Consider a networked Cournot game satisfying assumptions (a) and (b). Then,

	\begin{itemize}
		\item There exists an equilibrium $\left(q^{*}, f^{*}\right)$;
		\item Assume that assumption (c) also holds true and the price functions are affine with $P_{j}^{\prime}=-\beta_{j}<0$ for $1\le j \le m$ as well, then the game is potential with unique equilibrium given by
		\begin{equation} \label{eq:pot}
			\left(q^{*}, f^{*}\right)=\underset{q \geq 0,|f| \leq c}{\operatorname{argmax}} \left[w(f, q)-\frac{1}{2} \sum_{j=1}^m \beta_{j}\left(H^{\top} q^2\right)_{j} \right] \: .
		\end{equation}
	\end{itemize}
\end{thm}

\medskip

\begin{pf} $\:$

	\begin{itemize}
		\item By assumption (a), for all $1 \le j \le m$ it exists $\bar{r} \in (0,+\infty)$ such that $P_j(\bar{r})=0$ and because $P_j$ is a monotone decreasing function, we have that $P_j(r_j) \le 0 \: \forall r_j \ge \overbar{r}$. In our case this implies that, for a fixed $f$ such that $\abs{f}\le c$, the utility of the generic producer $i$ \eqref{eq1} becomes non-positive for $\left(Bf+H^{\top}q\right)_k \ge \bar{r} $ where $k$ is the only index such that $H_{ik}=1$ (assumption (b)). Notice that, in particular, the utility is certainly non-positive for $q_i \ge \bar{r}+(Bc)_k$. This implies that we can effectively bound the action of each producers $i$ such that $0 \le q_i \le \bar{r}+(Bc)_k$ as the previous considerations guarantee that
		$$ \underset{0 \le q_i \le \bar{r}+(Bc)_k}{\operatorname{argmax}} u_{i}\left(q_{i}, q_{-i}, f\right) = \underset{q_i \ge 0}{\operatorname{argmax}} \:  u_{i}\left(q_{i}, q_{-i}, f\right) \: .$$
		
		With this in mind, we can notice that the strategy sets are non-empty, convex and compact for each player (both the producers and the market maker). Under assumptions (a) and (b), we have that $u_i$ and $w$ are continuous for all $i\in \{1,\dots,n\}$; moreover, for all $q_{-i}$ and $f$ such that $\abs{f} \le c$ we have that $q_i \mapsto u_i(q_i,q_{-i},f)$ is concave and for all $q$ we have that $f\mapsto w(f,q)$ is also concave. Hence, by Theorem 1.2 in \cite{Dutang2013} it exists a Nash equilibrium.
		\item We need to prove that the function $\Phi(f,q) = w(f, q)-\frac{1}{2} \sum_{j=1}^m \beta_{j}\left(H^{\top} q^2\right)_{j} $ is a potential. Notice that $w(f,q)-\Phi(f,q)$ does not depend on $f$. Hence, for every $q \in \R_+^n$ and for each feasible $f_1,f_2$ we have 
		\begin{equation}
			\Phi(f_1,q)-\Phi(f_2,q) = 	w(f_1,q)-w(f_2,q) \: .
		\end{equation}
		To finish the proof, differentiate $\Phi(f,q)$ with respect to $q_i$:
		\begin{align*}
			\frac{\partial}{\partial q_i}\Phi(f,q) &=  \sum_{j=1}^m H_{i j} P_{j}\left(\left(B f+H^{\top} q\right)_{j}\right)\\[2ex] 
			& -\cost_{i}^{\prime}\left(q_{i}\right)-q_i\sum_{j=1}^m \beta_j H_{ij} \\[2ex]
			&= 	\frac{\partial}{\partial q_i} \left[q_i \sum_{j=1}^m H_{i j} P_{j}\left(\left(B f+H^{\top} q\right)_{j}\right)-\cost_{i} \right] \\[2ex] 
			&=\frac{\partial}{\partial q_i} u_i(f,q) \: .
		\end{align*}
		
	\end{itemize}
	
	Where we used the fact that $\frac{\partial}{\partial q_i} H_{i j} P_{j}\left(\left(B f+H^{\top} q\right)_{j}\right) = H_{ij}^2 P_j^{\prime}\left(\left(B f+H^{\top} q\right)_{j}\right) = -\beta_j  H_{ij} $.

\end{pf}

The following result establishes a connection between price differences at equilibrium and capacity bottlenecks in the power network and it is a direct application of the very general result of Theorem \ref{th2}.

\smallskip

\begin{corollary}
	\label{col1}
	Consider a networked Cournot game satisfying assumptions (a), (b) and (c) and an equilibrium $ (f^*, q^*) $ with prices $p_j^*=P_j\left(Bf^*+H^{\top}q^*\right)_j$. Then, if there exist $i,j$ such that $p_i^* < p_j^*$ then 
	
	\begin{enumerate}
		\item It exists a saturated $i-j$ cut.
		\item There is no flow on any $j-i$ path.
	\end{enumerate}
\end{corollary}

\smallskip

\begin{pf}
	It follows immediately from Theorem \ref{th2} by noticing that the best response of the market maker coincides with \eqref{t1} and in this case $\frac{\partial v}{\partial r_i} =p_i$ as it can be seen by deriving \eqref{eqmm}, hence condition \eqref{t2} reads as  $p_i^* < p_j^*$.
	
\end{pf}

Notice that thanks to the generality of Theorem \ref{th2}, Corollary \ref{col1} would still hold true even if we were to drop assumptions (a) and (b). Corollary \ref{col1} formally proves
the existence of a link between capacity bottlenecks and
price differences, a very well known phenomenon in real-world power networks. We show this effect in the following example where for sake of simplicity we consider a model satisfying all assumptions (a), (b) and (c).

\medskip


\subsubsection{Example.}

	We consider a simplified model of the Italian power network shown in Fig. \ref{fig3} consisting of 22 markets (the nodes of the network) present in different regions of the country. The blue nodes indicate the three main hubs (north, central north and central south parts of Italy). The topology of the network and the capacities of power lines are publicly available at \cite{org}. We consider the market maker utility $w$ to be equal to the Marshallian welfare \eqref{eq2}. For sake of simplicity, we consider the same affine price function for all markets and the same quadratic cost function for all producers: $P_j=\alpha-\beta d_j$ for $1 \le j \le m$ (measured in Euros (\euro) per Mega Watts-hour (MWh)) and $\cost_i=\theta q_i^2$ for $1 \le i \le n$ (measured in Euros). Although these assumptions are of course not realistic for a real power network, it will help us isolate the specific effect of capacity bottlenecks on price differences without mixing it up with other effects due to a mismatch between the parameters characterizing the utilities of the producers. 
	We consider a total of 31 producers that supply energy to the network and the number of producers for each market is indicated by the red digit next to each market. 
	The capacity for each line is indicated as the weight of the corresponding link and is measured in Mega Watts (MW). For the demand and cost functions, we choose the following values: $\alpha =120 \frac{\text{\euro}}{\text{MWh}}, \beta = 0.04 \frac{\text{\euro}}{\text{MWh}^2} $ and $\theta=0.01 \frac{\text{\euro}}{\text{MWh}^2}$.
	
	\begin{tikzpicture}[thick, x=0.75pt,y=0.75pt,yscale=-1,xscale=1]
		
		\draw   (200,40) .. controls (200,34.48) and (204.48,30) .. (210,30) .. controls (215.52,30) and (220,34.48) .. (220,40) .. controls (220,45.52) and (215.52,50) .. (210,50) .. controls (204.48,50) and (200,45.52) .. (200,40) -- cycle ;
		\draw  [color={rgb, 255:red, 38; green, 92; blue, 239 }  ,draw opacity=1 ][line width=1.5]  (210,90) .. controls (210,84.48) and (214.48,80) .. (220,80) .. controls (225.52,80) and (230,84.48) .. (230,90) .. controls (230,95.52) and (225.52,100) .. (220,100) .. controls (214.48,100) and (210,95.52) .. (210,90) -- cycle ;
		\draw   (260,60) .. controls (260,54.48) and (264.48,50) .. (270,50) .. controls (275.52,50) and (280,54.48) .. (280,60) .. controls (280,65.52) and (275.52,70) .. (270,70) .. controls (264.48,70) and (260,65.52) .. (260,60) -- cycle ;
		\draw   (160,90) .. controls (160,84.48) and (164.48,80) .. (170,80) .. controls (175.52,80) and (180,84.48) .. (180,90) .. controls (180,95.52) and (175.52,100) .. (170,100) .. controls (164.48,100) and (160,95.52) .. (160,90) -- cycle ;
		\draw   (330,40) .. controls (330,34.48) and (325.52,30) .. (320,30) .. controls (314.48,30) and (310,34.48) .. (310,40) .. controls (310,45.52) and (314.48,50) .. (320,50) .. controls (325.52,50) and (330,45.52) .. (330,40) -- cycle ;
		\draw   (130,50) .. controls (130,44.48) and (134.48,40) .. (140,40) .. controls (145.52,40) and (150,44.48) .. (150,50) .. controls (150,55.52) and (145.52,60) .. (140,60) .. controls (134.48,60) and (130,55.52) .. (130,50) -- cycle ;
		\draw   (150,140) .. controls (150,134.48) and (154.48,130) .. (160,130) .. controls (165.52,130) and (170,134.48) .. (170,140) .. controls (170,145.52) and (165.52,150) .. (160,150) .. controls (154.48,150) and (150,145.52) .. (150,140) -- cycle ;
		\draw   (270,110) .. controls (270,104.48) and (274.48,100) .. (280,100) .. controls (285.52,100) and (290,104.48) .. (290,110) .. controls (290,115.52) and (285.52,120) .. (280,120) .. controls (274.48,120) and (270,115.52) .. (270,110) -- cycle ;
		\draw   (320,130) .. controls (320,124.48) and (324.48,120) .. (330,120) .. controls (335.52,120) and (340,124.48) .. (340,130) .. controls (340,135.52) and (335.52,140) .. (330,140) .. controls (324.48,140) and (320,135.52) .. (320,130) -- cycle ;
		\draw   (190,160) .. controls (190,154.48) and (194.48,150) .. (200,150) .. controls (205.52,150) and (210,154.48) .. (210,160) .. controls (210,165.52) and (205.52,170) .. (200,170) .. controls (194.48,170) and (190,165.52) .. (190,160) -- cycle ;
		\draw  [color={rgb, 255:red, 37; green, 82; blue, 232 }  ,draw opacity=1 ][line width=1.5]  (240,140) .. controls (240,134.48) and (244.48,130) .. (250,130) .. controls (255.52,130) and (260,134.48) .. (260,140) .. controls (260,145.52) and (255.52,150) .. (250,150) .. controls (244.48,150) and (240,145.52) .. (240,140) -- cycle ;
		\draw  [color={rgb, 255:red, 32; green, 86; blue, 232 }  ,draw opacity=1 ][line width=1.5]  (290,170) .. controls (290,164.48) and (294.48,160) .. (300,160) .. controls (305.52,160) and (310,164.48) .. (310,170) .. controls (310,175.52) and (305.52,180) .. (300,180) .. controls (294.48,180) and (290,175.52) .. (290,170) -- cycle ;
		\draw   (230,190) .. controls (230,184.48) and (234.48,180) .. (240,180) .. controls (245.52,180) and (250,184.48) .. (250,190) .. controls (250,195.52) and (245.52,200) .. (240,200) .. controls (234.48,200) and (230,195.52) .. (230,190) -- cycle ;
		\draw   (160,200) .. controls (160,194.48) and (164.48,190) .. (170,190) .. controls (175.52,190) and (180,194.48) .. (180,200) .. controls (180,205.52) and (175.52,210) .. (170,210) .. controls (164.48,210) and (160,205.52) .. (160,200) -- cycle ;
		\draw    (146,58.67) -- (164,81.67) ;
		\draw    (180,90) -- (210,90) ;
		\draw    (210,50) -- (220,80) ;
		\draw    (262,64.67) -- (230,90) ;
		\draw    (312,45.67) -- (280,60) ;
		\draw    (230,90) -- (270,110) ;
		\draw    (288,116.67) -- (320,130) ;
		\draw    (220,100) -- (167.5,134) ;
		\draw    (220,100) -- (244.5,131) ;
		\draw    (240.5,145) -- (210,160) ;
		\draw    (259.5,145) -- (290,170) ;
		\draw    (210,160) -- (230,190) ;
		\draw    (250,190) -- (290,170) ;
		\draw    (180,200) -- (230,190) ;
		\draw   (333,179) .. controls (333,173.48) and (337.48,169) .. (343,169) .. controls (348.52,169) and (353,173.48) .. (353,179) .. controls (353,184.52) and (348.52,189) .. (343,189) .. controls (337.48,189) and (333,184.52) .. (333,179) -- cycle ;
		\draw   (377,163) .. controls (377,157.48) and (381.48,153) .. (387,153) .. controls (392.52,153) and (397,157.48) .. (397,163) .. controls (397,168.52) and (392.52,173) .. (387,173) .. controls (381.48,173) and (377,168.52) .. (377,163) -- cycle ;
		\draw   (385,193) .. controls (385,187.48) and (389.48,183) .. (395,183) .. controls (400.52,183) and (405,187.48) .. (405,193) .. controls (405,198.52) and (400.52,203) .. (395,203) .. controls (389.48,203) and (385,198.52) .. (385,193) -- cycle ;
		\draw   (417,186) .. controls (417,180.48) and (421.48,176) .. (427,176) .. controls (432.52,176) and (437,180.48) .. (437,186) .. controls (437,191.52) and (432.52,196) .. (427,196) .. controls (421.48,196) and (417,191.52) .. (417,186) -- cycle ;
		\draw   (318,221) .. controls (318,215.48) and (322.48,211) .. (328,211) .. controls (333.52,211) and (338,215.48) .. (338,221) .. controls (338,226.52) and (333.52,231) .. (328,231) .. controls (322.48,231) and (318,226.52) .. (318,221) -- cycle ;
		\draw   (268,232) .. controls (268,226.48) and (272.48,222) .. (278,222) .. controls (283.52,222) and (288,226.48) .. (288,232) .. controls (288,237.52) and (283.52,242) .. (278,242) .. controls (272.48,242) and (268,237.52) .. (268,232) -- cycle ;
		\draw   (229,260) .. controls (229,254.48) and (233.48,250) .. (239,250) .. controls (244.52,250) and (249,254.48) .. (249,260) .. controls (249,265.52) and (244.52,270) .. (239,270) .. controls (233.48,270) and (229,265.52) .. (229,260) -- cycle ;
		\draw   (292,261) .. controls (292,255.48) and (296.48,251) .. (302,251) .. controls (307.52,251) and (312,255.48) .. (312,261) .. controls (312,266.52) and (307.52,271) .. (302,271) .. controls (296.48,271) and (292,266.52) .. (292,261) -- cycle ;
		\draw    (333,179) -- (310,170) ;
		\draw    (377,163) -- (353,179) ;
		\draw    (353,179) -- (385,193) ;
		\draw    (405,193) -- (417,186.87) ;
		\draw    (328,211) -- (343,189) ;
		\draw    (288,232) -- (318,221) ;
		\draw    (239,250) -- (268,232) ;
		\draw    (302,251) -- (288,232) ;
		
		\draw (181,65) node [anchor=north west][inner sep=0.75pt]  [font=\footnotesize,color={rgb, 255:red, 26; green, 65; blue, 238 }  ,opacity=1 ] [align=left] {North};
		\draw (263,131) node [anchor=north west][inner sep=0.75pt]  [font=\footnotesize,color={rgb, 255:red, 26; green, 65; blue, 238 }  ,opacity=1 ] [align=left] {CNorth};
		\draw (310,149) node [anchor=north west][inner sep=0.75pt]  [font=\footnotesize,color={rgb, 255:red, 26; green, 65; blue, 238 }  ,opacity=1 ] [align=left] {CSouth};
		\draw (130,72.4) node [anchor=north west][inner sep=0.75pt]  [font=\tiny]  {$1300$};
		\draw (181,82.4) node [anchor=north west][inner sep=0.75pt]  [font=\tiny]  {$10000$};
		\draw (181,52.4) node [anchor=north west][inner sep=0.75pt]  [font=\tiny]  {$10000$};
		\draw (231,62.4) node [anchor=north west][inner sep=0.75pt]  [font=\tiny]  {$10000$};
		\draw (284,42.4) node [anchor=north west][inner sep=0.75pt]  [font=\tiny]  {$200$};
		\draw (246,92.4) node [anchor=north west][inner sep=0.75pt]  [font=\tiny]  {$10000$};
		\draw (304,112.4) node [anchor=north west][inner sep=0.75pt]  [font=\tiny]  {$640$};
		\draw (211,112.4) node [anchor=north west][inner sep=0.75pt]  [font=\tiny]  {$2800$};
		\draw (161,112.4) node [anchor=north west][inner sep=0.75pt]  [font=\tiny]  {$10000$};
		\draw (214,142.4) node [anchor=north west][inner sep=0.75pt]  [font=\tiny]  {$300$};
		\draw (221,162.4) node [anchor=north west][inner sep=0.75pt]  [font=\tiny]  {$300$};
		\draw (194,202.4) node [anchor=north west][inner sep=0.75pt]  [font=\tiny]  {$100$};
		\draw (254,172.4) node [anchor=north west][inner sep=0.75pt]  [font=\tiny]  {$800$};
		\draw (311,182.4) node [anchor=north west][inner sep=0.75pt]  [font=\tiny]  {$8000$};
		\draw (350,162.4) node [anchor=north west][inner sep=0.75pt]  [font=\tiny]  {$8000$};
		\draw (406,202.4) node [anchor=north west][inner sep=0.75pt]  [font=\tiny]  {$10000$};
		\draw (360,192.4) node [anchor=north west][inner sep=0.75pt]  [font=\tiny]  {$8000$};
		\draw (337.5,203.4) node [anchor=north west][inner sep=0.75pt]  [font=\tiny]  {$8000$};
		\draw (291,212.4) node [anchor=north west][inner sep=0.75pt]  [font=\tiny]  {$1200$};
		\draw (241,232.4) node [anchor=north west][inner sep=0.75pt]  [font=\tiny]  {$200$};
		\draw (301,242.4) node [anchor=north west][inner sep=0.75pt]  [font=\tiny]  {$8000$};
		\draw (311,69) node [anchor=north west][inner sep=0.75pt]  [font=\scriptsize] [align=left] {\mbox{-} Line capacities in MW\\\textcolor[rgb]{0.82,0.01,0.11}{- Number of producers per node}};
		\draw (252,153.4) node [anchor=north west][inner sep=0.75pt]  [font=\tiny]  {$2700$};
		\draw (137,48.4) node [anchor=north west][inner sep=0.75pt]  [font=\tiny]  {$1$};
		\draw (166,87.07) node [anchor=north west][inner sep=0.75pt]  [font=\tiny]  {$2$};
		\draw (206,38.4) node [anchor=north west][inner sep=0.75pt]  [font=\tiny]  {$3$};
		\draw (216,88.4) node [anchor=north west][inner sep=0.75pt]  [font=\tiny]  {$4$};
		\draw (267,58.4) node [anchor=north west][inner sep=0.75pt]  [font=\tiny]  {$5$};
		\draw (316,38.4) node [anchor=north west][inner sep=0.75pt]  [font=\tiny]  {$6$};
		\draw (157,138.4) node [anchor=north west][inner sep=0.75pt]  [font=\tiny]  {$7$};
		\draw (324,128.4) node [anchor=north west][inner sep=0.75pt]  [font=\tiny]  {$10$};
		\draw (246,137.4) node [anchor=north west][inner sep=0.75pt]  [font=\tiny]  {$8$};
		\draw (276,108.4) node [anchor=north west][inner sep=0.75pt]  [font=\tiny]  {$9$};
		\draw (165,198.4) node [anchor=north west][inner sep=0.75pt]  [font=\tiny]  {$11$};
		\draw (195,158.4) node [anchor=north west][inner sep=0.75pt]  [font=\tiny]  {$12$};
		\draw (296,168.4) node [anchor=north west][inner sep=0.75pt]  [font=\tiny]  {$13$};
		\draw (235,187.4) node [anchor=north west][inner sep=0.75pt]  [font=\tiny]  {$15$};
		\draw (337,177.4) node [anchor=north west][inner sep=0.75pt]  [font=\tiny]  {$16$};
		\draw (382,159.4) node [anchor=north west][inner sep=0.75pt]  [font=\tiny]  {$14$};
		\draw (390,191.4) node [anchor=north west][inner sep=0.75pt]  [font=\tiny]  {$17$};
		\draw (421,184.4) node [anchor=north west][inner sep=0.75pt]  [font=\tiny]  {$18$};
		\draw (322,217.4) node [anchor=north west][inner sep=0.75pt]  [font=\tiny]  {$19$};
		\draw (273,229.4) node [anchor=north west][inner sep=0.75pt]  [font=\tiny]  {$20$};
		\draw (233,257.4) node [anchor=north west][inner sep=0.75pt]  [font=\tiny]  {$21$};
		\draw (297,260.4) node [anchor=north west][inner sep=0.75pt]  [font=\tiny]  {$22$};
		\draw (140,31.4) node [anchor=north west][inner sep=0.75pt]  [font=\tiny,color={rgb, 255:red, 208; green, 2; blue, 27 }  ,opacity=1 ]  {$0$};
		\draw (152,92.4) node [anchor=north west][inner sep=0.75pt]  [font=\tiny,color={rgb, 255:red, 208; green, 2; blue, 27 }  ,opacity=1 ]  {$2$};
		\draw (192,26.4) node [anchor=north west][inner sep=0.75pt]  [font=\tiny,color={rgb, 255:red, 208; green, 2; blue, 27 }  ,opacity=1 ]  {$3$};
		\draw (226,73.4) node [anchor=north west][inner sep=0.75pt]  [font=\tiny,color={rgb, 255:red, 208; green, 2; blue, 27 }  ,opacity=1 ]  {$1$};
		\draw (262,42.4) node [anchor=north west][inner sep=0.75pt]  [font=\tiny,color={rgb, 255:red, 208; green, 2; blue, 27 }  ,opacity=1 ]  {$0$};
		\draw (319,22.4) node [anchor=north west][inner sep=0.75pt]  [font=\tiny,color={rgb, 255:red, 208; green, 2; blue, 27 }  ,opacity=1 ]  {$1$};
		\draw (154,122.4) node [anchor=north west][inner sep=0.75pt]  [font=\tiny,color={rgb, 255:red, 208; green, 2; blue, 27 }  ,opacity=1 ]  {$2$};
		\draw (245,119.4) node [anchor=north west][inner sep=0.75pt]  [font=\tiny,color={rgb, 255:red, 208; green, 2; blue, 27 }  ,opacity=1 ]  {$1$};
		\draw (280,88.4) node [anchor=north west][inner sep=0.75pt]  [font=\tiny,color={rgb, 255:red, 208; green, 2; blue, 27 }  ,opacity=1 ]  {$0$};
		\draw (331,108.4) node [anchor=north west][inner sep=0.75pt]  [font=\tiny,color={rgb, 255:red, 208; green, 2; blue, 27 }  ,opacity=1 ]  {$2$};
		\draw (165,180.4) node [anchor=north west][inner sep=0.75pt]  [font=\tiny,color={rgb, 255:red, 208; green, 2; blue, 27 }  ,opacity=1 ]  {$2$};
		\draw (196,141.4) node [anchor=north west][inner sep=0.75pt]  [font=\tiny,color={rgb, 255:red, 208; green, 2; blue, 27 }  ,opacity=1 ]  {$1$};
		\draw (293,152.4) node [anchor=north west][inner sep=0.75pt]  [font=\tiny,color={rgb, 255:red, 208; green, 2; blue, 27 }  ,opacity=1 ]  {$1$};
		\draw (383,141.4) node [anchor=north west][inner sep=0.75pt]  [font=\tiny,color={rgb, 255:red, 208; green, 2; blue, 27 }  ,opacity=1 ]  {$1$};
		\draw (237,202.4) node [anchor=north west][inner sep=0.75pt]  [font=\tiny,color={rgb, 255:red, 208; green, 2; blue, 27 }  ,opacity=1 ]  {$5$};
		\draw (346,190.4) node [anchor=north west][inner sep=0.75pt]  [font=\tiny,color={rgb, 255:red, 208; green, 2; blue, 27 }  ,opacity=1 ]  {$2$};
		\draw (391,206.4) node [anchor=north west][inner sep=0.75pt]  [font=\tiny,color={rgb, 255:red, 208; green, 2; blue, 27 }  ,opacity=1 ]  {$1$};
		\draw (425,166.4) node [anchor=north west][inner sep=0.75pt]  [font=\tiny,color={rgb, 255:red, 208; green, 2; blue, 27 }  ,opacity=1 ]  {$2$};
		\draw (341,223.4) node [anchor=north west][inner sep=0.75pt]  [font=\tiny,color={rgb, 255:red, 208; green, 2; blue, 27 }  ,opacity=1 ]  {$2$};
		\draw (270,212.4) node [anchor=north west][inner sep=0.75pt]  [font=\tiny,color={rgb, 255:red, 208; green, 2; blue, 27 }  ,opacity=1 ]  {$0$};
		\draw (218,259.4) node [anchor=north west][inner sep=0.75pt]  [font=\tiny,color={rgb, 255:red, 208; green, 2; blue, 27 }  ,opacity=1 ]  {$1$};
		\draw (281,259.4) node [anchor=north west][inner sep=0.75pt]  [font=\tiny,color={rgb, 255:red, 208; green, 2; blue, 27 }  ,opacity=1 ]  {$1$};

	\end{tikzpicture}

	
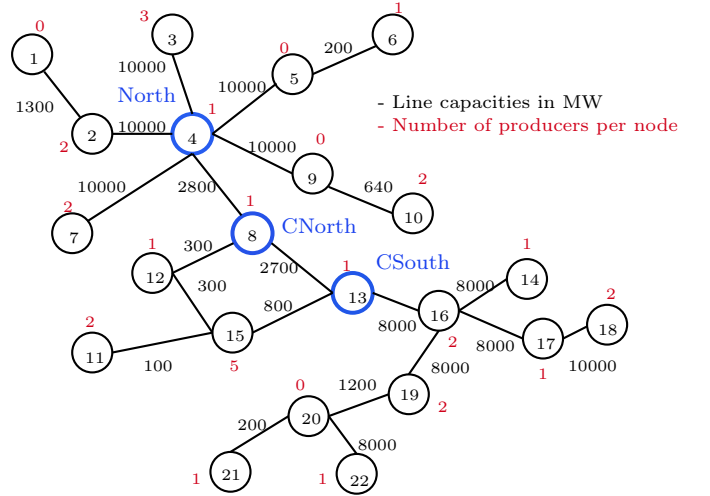
\captionof{figure}{{Italian power network with line capacities measured in MW.}}
	\label{fig3}
	
	\bigskip	
	
	\medskip
	
	Notice that assumptions (a), (b) and (c) are satisfied and the price functions are affine; by Theorem \ref{th1} this means that we have a potential game with a unique Nash equilibria that can be found solving \eqref{eq:pot}. By doing that, we find the equilibrium prices as shown in Tab. \ref{tab1} and in Fig. \ref{fig4}.
	
	\medskip
	
	\begin{tikzpicture}[thick, x=0.75pt,y=0.75pt,yscale=-1,xscale=1]
		
		
		\draw  [color={rgb, 255:red, 245; green, 166; blue, 35 }  ,draw opacity=1 ] (220,43) .. controls (220,37.48) and (224.48,33) .. (230,33) .. controls (235.52,33) and (240,37.48) .. (240,43) .. controls (240,48.52) and (235.52,53) .. (230,53) .. controls (224.48,53) and (220,48.52) .. (220,43) -- cycle ;
		\draw  [color={rgb, 255:red, 245; green, 166; blue, 35 }  ,draw opacity=1 ][line width=1.5]  (230,93) .. controls (230,87.48) and (234.48,83) .. (240,83) .. controls (245.52,83) and (250,87.48) .. (250,93) .. controls (250,98.52) and (245.52,103) .. (240,103) .. controls (234.48,103) and (230,98.52) .. (230,93) -- cycle ;
		\draw  [color={rgb, 255:red, 245; green, 166; blue, 35 }  ,draw opacity=1 ] (280,63) .. controls (280,57.48) and (284.48,53) .. (290,53) .. controls (295.52,53) and (300,57.48) .. (300,63) .. controls (300,68.52) and (295.52,73) .. (290,73) .. controls (284.48,73) and (280,68.52) .. (280,63) -- cycle ;
		\draw  [color={rgb, 255:red, 245; green, 166; blue, 35 }  ,draw opacity=1 ] (180,93) .. controls (180,87.48) and (184.48,83) .. (190,83) .. controls (195.52,83) and (200,87.48) .. (200,93) .. controls (200,98.52) and (195.52,103) .. (190,103) .. controls (184.48,103) and (180,98.52) .. (180,93) -- cycle ;
		\draw  [color={rgb, 255:red, 208; green, 2; blue, 27 }  ,draw opacity=1 ] (350,43) .. controls (350,37.48) and (345.52,33) .. (340,33) .. controls (334.48,33) and (330,37.48) .. (330,43) .. controls (330,48.52) and (334.48,53) .. (340,53) .. controls (345.52,53) and (350,48.52) .. (350,43) -- cycle ;
		\draw  [color={rgb, 255:red, 208; green, 2; blue, 27 }  ,draw opacity=1 ] (150,53) .. controls (150,47.48) and (154.48,43) .. (160,43) .. controls (165.52,43) and (170,47.48) .. (170,53) .. controls (170,58.52) and (165.52,63) .. (160,63) .. controls (154.48,63) and (150,58.52) .. (150,53) -- cycle ;
		\draw  [color={rgb, 255:red, 245; green, 166; blue, 35 }  ,draw opacity=1 ] (170,143) .. controls (170,137.48) and (174.48,133) .. (180,133) .. controls (185.52,133) and (190,137.48) .. (190,143) .. controls (190,148.52) and (185.52,153) .. (180,153) .. controls (174.48,153) and (170,148.52) .. (170,143) -- cycle ;
		\draw  [color={rgb, 255:red, 245; green, 166; blue, 35 }  ,draw opacity=1 ] (290,113) .. controls (290,107.48) and (294.48,103) .. (300,103) .. controls (305.52,103) and (310,107.48) .. (310,113) .. controls (310,118.52) and (305.52,123) .. (300,123) .. controls (294.48,123) and (290,118.52) .. (290,113) -- cycle ;
		\draw  [color={rgb, 255:red, 248; green, 231; blue, 28 }  ,draw opacity=1 ] (340,133) .. controls (340,127.48) and (344.48,123) .. (350,123) .. controls (355.52,123) and (360,127.48) .. (360,133) .. controls (360,138.52) and (355.52,143) .. (350,143) .. controls (344.48,143) and (340,138.52) .. (340,133) -- cycle ;
		\draw  [color={rgb, 255:red, 65; green, 117; blue, 5 }  ,draw opacity=1 ] (210,163) .. controls (210,157.48) and (214.48,153) .. (220,153) .. controls (225.52,153) and (230,157.48) .. (230,163) .. controls (230,168.52) and (225.52,173) .. (220,173) .. controls (214.48,173) and (210,168.52) .. (210,163) -- cycle ;
		\draw  [color={rgb, 255:red, 245; green, 166; blue, 35 }  ,draw opacity=1 ][line width=1.5]  (260,143) .. controls (260,137.48) and (264.48,133) .. (270,133) .. controls (275.52,133) and (280,137.48) .. (280,143) .. controls (280,148.52) and (275.52,153) .. (270,153) .. controls (264.48,153) and (260,148.52) .. (260,143) -- cycle ;
		\draw  [color={rgb, 255:red, 245; green, 166; blue, 35 }  ,draw opacity=1 ][line width=1.5]  (310,173) .. controls (310,167.48) and (314.48,163) .. (320,163) .. controls (325.52,163) and (330,167.48) .. (330,173) .. controls (330,178.52) and (325.52,183) .. (320,183) .. controls (314.48,183) and (310,178.52) .. (310,173) -- cycle ;
		\draw  [color={rgb, 255:red, 65; green, 117; blue, 5 }  ,draw opacity=1 ] (250,193) .. controls (250,187.48) and (254.48,183) .. (260,183) .. controls (265.52,183) and (270,187.48) .. (270,193) .. controls (270,198.52) and (265.52,203) .. (260,203) .. controls (254.48,203) and (250,198.52) .. (250,193) -- cycle ;
		\draw  [color={rgb, 255:red, 126; green, 211; blue, 33 }  ,draw opacity=1 ] (180,203) .. controls (180,197.48) and (184.48,193) .. (190,193) .. controls (195.52,193) and (200,197.48) .. (200,203) .. controls (200,208.52) and (195.52,213) .. (190,213) .. controls (184.48,213) and (180,208.52) .. (180,203) -- cycle ;
		\draw    (167.85,64.03) -- (172.78,70.33) .. controls (175.12,70.62) and (176.15,71.93) .. (175.86,74.27) .. controls (175.57,76.61) and (176.6,77.92) .. (178.94,78.2) .. controls (181.28,78.49) and (182.31,79.8) .. (182.02,82.14) -- (184,84.67) -- (184,84.67) ;
		\draw [shift={(166,61.67)}, rotate = 51.95] [fill={rgb, 255:red, 0; green, 0; blue, 0 }  ][line width=0.08]  [draw opacity=0] (8.93,-4.29) -- (0,0) -- (8.93,4.29) -- cycle    ;
		\draw    (203,93) -- (230,93) ;
		\draw [shift={(200,93)}, rotate = 0] [fill={rgb, 255:red, 0; green, 0; blue, 0 }  ][line width=0.08]  [draw opacity=0] (8.93,-4.29) -- (0,0) -- (8.93,4.29) -- cycle    ;
		\draw    (230,53) -- (239.05,80.15) ;
		\draw [shift={(240,83)}, rotate = 251.57] [fill={rgb, 255:red, 0; green, 0; blue, 0 }  ][line width=0.08]  [draw opacity=0] (8.93,-4.29) -- (0,0) -- (8.93,4.29) -- cycle    ;
		\draw    (279.65,69.53) -- (250,93) ;
		\draw [shift={(282,67.67)}, rotate = 141.63] [fill={rgb, 255:red, 0; green, 0; blue, 0 }  ][line width=0.08]  [draw opacity=0] (8.93,-4.29) -- (0,0) -- (8.93,4.29) -- cycle    ;
		\draw    (329.26,49.89) -- (321.96,53.16) .. controls (321.12,55.37) and (319.6,56.05) .. (317.4,55.21) .. controls (315.2,54.37) and (313.68,55.05) .. (312.83,57.25) .. controls (312,59.46) and (310.48,60.14) .. (308.27,59.3) .. controls (306.07,58.46) and (304.55,59.14) .. (303.71,61.34) -- (300,63) -- (300,63) ;
		\draw [shift={(332,48.67)}, rotate = 155.87] [fill={rgb, 255:red, 0; green, 0; blue, 0 }  ][line width=0.08]  [draw opacity=0] (8.93,-4.29) -- (0,0) -- (8.93,4.29) -- cycle    ;
		\draw    (250,93) -- (290,113) ;
		\draw    (310.77,120.82) -- (318.15,123.9) .. controls (320.33,123) and (321.87,123.64) .. (322.77,125.82) .. controls (323.66,128) and (325.2,128.64) .. (327.38,127.74) .. controls (329.56,126.85) and (331.1,127.49) .. (332,129.67) .. controls (332.9,131.85) and (334.44,132.49) .. (336.62,131.59) -- (340,133) -- (340,133) ;
		\draw [shift={(308,119.67)}, rotate = 22.62] [fill={rgb, 255:red, 0; green, 0; blue, 0 }  ][line width=0.08]  [draw opacity=0] (8.93,-4.29) -- (0,0) -- (8.93,4.29) -- cycle    ;
		\draw    (237.48,104.63) -- (187.5,137) ;
		\draw [shift={(240,103)}, rotate = 147.07] [fill={rgb, 255:red, 0; green, 0; blue, 0 }  ][line width=0.08]  [draw opacity=0] (8.93,-4.29) -- (0,0) -- (8.93,4.29) -- cycle    ;
		\draw    (241.86,105.35) -- (264.5,134) ;
		\draw [shift={(240,103)}, rotate = 51.68] [fill={rgb, 255:red, 0; green, 0; blue, 0 }  ][line width=0.08]  [draw opacity=0] (8.93,-4.29) -- (0,0) -- (8.93,4.29) -- cycle    ;
		\draw    (257.81,149.32) -- (250.63,152.85) .. controls (249.87,155.08) and (248.37,155.82) .. (246.14,155.06) .. controls (243.91,154.3) and (242.41,155.04) .. (241.66,157.27) .. controls (240.89,159.5) and (239.4,160.23) .. (237.17,159.47) .. controls (234.94,158.71) and (233.44,159.45) .. (232.68,161.68) -- (230,163) -- (230,163) ;
		\draw [shift={(260.5,148)}, rotate = 153.81] [fill={rgb, 255:red, 0; green, 0; blue, 0 }  ][line width=0.08]  [draw opacity=0] (8.93,-4.29) -- (0,0) -- (8.93,4.29) -- cycle    ;
		\draw    (281.82,149.9) -- (310,173) ;
		\draw [shift={(279.5,148)}, rotate = 39.34] [fill={rgb, 255:red, 0; green, 0; blue, 0 }  ][line width=0.08]  [draw opacity=0] (8.93,-4.29) -- (0,0) -- (8.93,4.29) -- cycle    ;
		\draw    (231.66,165.5) -- (250,193) ;
		\draw [shift={(230,163)}, rotate = 56.31] [fill={rgb, 255:red, 0; green, 0; blue, 0 }  ][line width=0.08]  [draw opacity=0] (8.93,-4.29) -- (0,0) -- (8.93,4.29) -- cycle    ;
		\draw    (270,193) .. controls (270.74,190.76) and (272.23,190.01) .. (274.47,190.76) .. controls (276.7,191.51) and (278.19,190.76) .. (278.94,188.53) .. controls (279.69,186.29) and (281.18,185.54) .. (283.42,186.29) .. controls (285.65,187.04) and (287.14,186.29) .. (287.89,184.06) .. controls (288.63,181.82) and (290.12,181.07) .. (292.36,181.82) .. controls (294.6,182.57) and (296.09,181.82) .. (296.83,179.58) -- (300.16,177.92) -- (307.32,174.34) ;
		\draw [shift={(310,173)}, rotate = 513.4300000000001] [fill={rgb, 255:red, 0; green, 0; blue, 0 }  ][line width=0.08]  [draw opacity=0] (8.93,-4.29) -- (0,0) -- (8.93,4.29) -- cycle    ;
		\draw    (202.94,202.41) -- (210.79,200.84) .. controls (212.1,198.88) and (213.73,198.55) .. (215.69,199.86) .. controls (217.65,201.17) and (219.28,200.84) .. (220.59,198.88) .. controls (221.9,196.92) and (223.54,196.59) .. (225.5,197.9) .. controls (227.46,199.21) and (229.09,198.88) .. (230.4,196.92) .. controls (231.71,194.96) and (233.34,194.63) .. (235.3,195.94) .. controls (237.26,197.25) and (238.89,196.92) .. (240.2,194.96) .. controls (241.51,193) and (243.15,192.67) .. (245.11,193.98) -- (250,193) -- (250,193) ;
		\draw [shift={(200,203)}, rotate = 348.69] [fill={rgb, 255:red, 0; green, 0; blue, 0 }  ][line width=0.08]  [draw opacity=0] (8.93,-4.29) -- (0,0) -- (8.93,4.29) -- cycle    ;
		\draw  [color={rgb, 255:red, 245; green, 166; blue, 35 }  ,draw opacity=1 ] (353,182) .. controls (353,176.48) and (357.48,172) .. (363,172) .. controls (368.52,172) and (373,176.48) .. (373,182) .. controls (373,187.52) and (368.52,192) .. (363,192) .. controls (357.48,192) and (353,187.52) .. (353,182) -- cycle ;
		\draw  [color={rgb, 255:red, 245; green, 166; blue, 35 }  ,draw opacity=1 ] (397,166) .. controls (397,160.48) and (401.48,156) .. (407,156) .. controls (412.52,156) and (417,160.48) .. (417,166) .. controls (417,171.52) and (412.52,176) .. (407,176) .. controls (401.48,176) and (397,171.52) .. (397,166) -- cycle ;
		\draw  [color={rgb, 255:red, 245; green, 166; blue, 35 }  ,draw opacity=1 ] (405,196) .. controls (405,190.48) and (409.48,186) .. (415,186) .. controls (420.52,186) and (425,190.48) .. (425,196) .. controls (425,201.52) and (420.52,206) .. (415,206) .. controls (409.48,206) and (405,201.52) .. (405,196) -- cycle ;
		\draw  [color={rgb, 255:red, 245; green, 166; blue, 35 }  ,draw opacity=1 ] (452,197) .. controls (452,191.48) and (456.48,187) .. (462,187) .. controls (467.52,187) and (472,191.48) .. (472,197) .. controls (472,202.52) and (467.52,207) .. (462,207) .. controls (456.48,207) and (452,202.52) .. (452,197) -- cycle ;
		\draw  [color={rgb, 255:red, 245; green, 166; blue, 35 }  ,draw opacity=1 ] (338,224) .. controls (338,218.48) and (342.48,214) .. (348,214) .. controls (353.52,214) and (358,218.48) .. (358,224) .. controls (358,229.52) and (353.52,234) .. (348,234) .. controls (342.48,234) and (338,229.52) .. (338,224) -- cycle ;
		\draw  [color={rgb, 255:red, 129; green, 3; blue, 3 }  ,draw opacity=1 ] (288,235) .. controls (288,229.48) and (292.48,225) .. (298,225) .. controls (303.52,225) and (308,229.48) .. (308,235) .. controls (308,240.52) and (303.52,245) .. (298,245) .. controls (292.48,245) and (288,240.52) .. (288,235) -- cycle ;
		\draw  [color={rgb, 255:red, 129; green, 3; blue, 3 }  ,draw opacity=1 ] (249,263) .. controls (249,257.48) and (253.48,253) .. (259,253) .. controls (264.52,253) and (269,257.48) .. (269,263) .. controls (269,268.52) and (264.52,273) .. (259,273) .. controls (253.48,273) and (249,268.52) .. (249,263) -- cycle ;
		\draw  [color={rgb, 255:red, 129; green, 3; blue, 3 }  ,draw opacity=1 ] (312,264) .. controls (312,258.48) and (316.48,254) .. (322,254) .. controls (327.52,254) and (332,258.48) .. (332,264) .. controls (332,269.52) and (327.52,274) .. (322,274) .. controls (316.48,274) and (312,269.52) .. (312,264) -- cycle ;
		\draw    (353,182) -- (332.79,174.09) ;
		\draw [shift={(330,173)}, rotate = 381.37] [fill={rgb, 255:red, 0; green, 0; blue, 0 }  ][line width=0.08]  [draw opacity=0] (8.93,-4.29) -- (0,0) -- (8.93,4.29) -- cycle    ;
		\draw    (394.5,167.66) -- (373,182) ;
		\draw [shift={(397,166)}, rotate = 146.31] [fill={rgb, 255:red, 0; green, 0; blue, 0 }  ][line width=0.08]  [draw opacity=0] (8.93,-4.29) -- (0,0) -- (8.93,4.29) -- cycle    ;
		\draw    (375.75,183.2) -- (405,196) ;
		\draw [shift={(373,182)}, rotate = 23.63] [fill={rgb, 255:red, 0; green, 0; blue, 0 }  ][line width=0.08]  [draw opacity=0] (8.93,-4.29) -- (0,0) -- (8.93,4.29) -- cycle    ;
		\draw    (428,196.11) -- (452,197) ;
		\draw [shift={(425,196)}, rotate = 2.12] [fill={rgb, 255:red, 0; green, 0; blue, 0 }  ][line width=0.08]  [draw opacity=0] (8.93,-4.29) -- (0,0) -- (8.93,4.29) -- cycle    ;
		\draw    (349.69,211.52) -- (363,192) ;
		\draw [shift={(348,214)}, rotate = 304.29] [fill={rgb, 255:red, 0; green, 0; blue, 0 }  ][line width=0.08]  [draw opacity=0] (8.93,-4.29) -- (0,0) -- (8.93,4.29) -- cycle    ;
		\draw    (310.82,233.97) -- (318.33,231.21) .. controls (319.32,229.07) and (320.88,228.5) .. (323.02,229.49) .. controls (325.16,230.48) and (326.73,229.91) .. (327.72,227.77) .. controls (328.71,225.63) and (330.27,225.06) .. (332.41,226.05) .. controls (334.55,227.04) and (336.12,226.47) .. (337.11,224.33) -- (338,224) -- (338,224) ;
		\draw [shift={(308,235)}, rotate = 339.86] [fill={rgb, 255:red, 0; green, 0; blue, 0 }  ][line width=0.08]  [draw opacity=0] (8.93,-4.29) -- (0,0) -- (8.93,4.29) -- cycle    ;
		\draw    (261.55,251.42) -- (288,235) ;
		\draw [shift={(259,253)}, rotate = 328.17] [fill={rgb, 255:red, 0; green, 0; blue, 0 }  ][line width=0.08]  [draw opacity=0] (8.93,-4.29) -- (0,0) -- (8.93,4.29) -- cycle    ;
		\draw    (320.22,251.58) -- (308,235) ;
		\draw [shift={(322,254)}, rotate = 233.62] [fill={rgb, 255:red, 0; green, 0; blue, 0 }  ][line width=0.08]  [draw opacity=0] (8.93,-4.29) -- (0,0) -- (8.93,4.29) -- cycle    ;
		
		\draw (201,68) node [anchor=north west][inner sep=0.75pt]  [font=\footnotesize,color={rgb, 255:red, 26; green, 65; blue, 238 }  ,opacity=1 ] [align=left] {North};
		\draw (283,134) node [anchor=north west][inner sep=0.75pt]  [font=\footnotesize,color={rgb, 255:red, 26; green, 65; blue, 238 }  ,opacity=1 ] [align=left] {CNorth};
		\draw (330,152) node [anchor=north west][inner sep=0.75pt]  [font=\footnotesize,color={rgb, 255:red, 26; green, 65; blue, 238 }  ,opacity=1 ] [align=left] {CSouth};
		\draw (150,75.4) node [anchor=north west][inner sep=0.75pt]  [font=\tiny]  {$1300$};
		\draw (208,82.4) node [anchor=north west][inner sep=0.75pt]  [font=\tiny]  {$570.4$};
		\draw (201,55.4) node [anchor=north west][inner sep=0.75pt]  [font=\tiny]  {$1795.2$};
		\draw (246,65.4) node [anchor=north west][inner sep=0.75pt]  [font=\tiny]  {$1601.6$};
		\draw (304,45.4) node [anchor=north west][inner sep=0.75pt]  [font=\tiny]  {$200$};
		\draw (267,92.4) node [anchor=north west][inner sep=0.75pt]  [font=\tiny]  {$761.6$};
		\draw (324,115.4) node [anchor=north west][inner sep=0.75pt]  [font=\tiny]  {$640$};
		\draw (231,122.4) node [anchor=north west][inner sep=0.75pt]  [font=\tiny]  {$744.8$};
		\draw (187,114.4) node [anchor=north west][inner sep=0.75pt]  [font=\tiny]  {$729.6$};
		\draw (234,145.4) node [anchor=north west][inner sep=0.75pt]  [font=\tiny]  {$300$};
		\draw (241,165.4) node [anchor=north west][inner sep=0.75pt]  [font=\tiny]  {$130$};
		\draw (214,205.4) node [anchor=north west][inner sep=0.75pt]  [font=\tiny]  {$100$};
		\draw (274,175.4) node [anchor=north west][inner sep=0.75pt]  [font=\tiny]  {$800$};
		\draw (331,185.4) node [anchor=north west][inner sep=0.75pt]  [font=\tiny]  {$316.8$};
		\draw (370,165.4) node [anchor=north west][inner sep=0.75pt]  [font=\tiny]  {$336$};
		\draw (426,205.4) node [anchor=north west][inner sep=0.75pt]  [font=\tiny]  {$729.6$};
		\draw (380,195.4) node [anchor=north west][inner sep=0.75pt]  [font=\tiny]  {$393.6$};
		\draw (357.5,206.4) node [anchor=north west][inner sep=0.75pt]  [font=\tiny]  {$470.4$};
		\draw (311,215.4) node [anchor=north west][inner sep=0.75pt]  [font=\tiny]  {$1200$};
		\draw (253,236.4) node [anchor=north west][inner sep=0.75pt]  [font=\tiny]  {$\approx 0$};
		\draw (320,241.4) node [anchor=north west][inner sep=0.75pt]  [font=\tiny]  {$\approx 0$};
		\draw (270,159.4) node [anchor=north west][inner sep=0.75pt]  [font=\tiny]  {$780.8$};
		\draw (157,51.4) node [anchor=north west][inner sep=0.75pt]  [font=\tiny]  {$1$};
		\draw (186,90.07) node [anchor=north west][inner sep=0.75pt]  [font=\tiny]  {$2$};
		\draw (226,41.4) node [anchor=north west][inner sep=0.75pt]  [font=\tiny]  {$3$};
		\draw (236,91.4) node [anchor=north west][inner sep=0.75pt]  [font=\tiny]  {$4$};
		\draw (287,61.4) node [anchor=north west][inner sep=0.75pt]  [font=\tiny]  {$5$};
		\draw (336,41.4) node [anchor=north west][inner sep=0.75pt]  [font=\tiny]  {$6$};
		\draw (177,141.4) node [anchor=north west][inner sep=0.75pt]  [font=\tiny]  {$7$};
		\draw (344,131.4) node [anchor=north west][inner sep=0.75pt]  [font=\tiny]  {$10$};
		\draw (266,140.4) node [anchor=north west][inner sep=0.75pt]  [font=\tiny]  {$8$};
		\draw (296,111.4) node [anchor=north west][inner sep=0.75pt]  [font=\tiny]  {$9$};
		\draw (185,201.4) node [anchor=north west][inner sep=0.75pt]  [font=\tiny]  {$11$};
		\draw (215,161.4) node [anchor=north west][inner sep=0.75pt]  [font=\tiny]  {$12$};
		\draw (316,171.4) node [anchor=north west][inner sep=0.75pt]  [font=\tiny]  {$13$};
		\draw (255,190.4) node [anchor=north west][inner sep=0.75pt]  [font=\tiny]  {$15$};
		\draw (357,180.4) node [anchor=north west][inner sep=0.75pt]  [font=\tiny]  {$16$};
		\draw (402,162.4) node [anchor=north west][inner sep=0.75pt]  [font=\tiny]  {$14$};
		\draw (410,194.4) node [anchor=north west][inner sep=0.75pt]  [font=\tiny]  {$17$};
		\draw (456,195.4) node [anchor=north west][inner sep=0.75pt]  [font=\tiny]  {$18$};
		\draw (342,220.4) node [anchor=north west][inner sep=0.75pt]  [font=\tiny]  {$19$};
		\draw (293,232.4) node [anchor=north west][inner sep=0.75pt]  [font=\tiny]  {$20$};
		\draw (253,260.4) node [anchor=north west][inner sep=0.75pt]  [font=\tiny]  {$21$};
		\draw (317,263.4) node [anchor=north west][inner sep=0.75pt]  [font=\tiny]  {$22$};
		\draw (153,34.4) node [anchor=north west][inner sep=0.75pt]  [font=\tiny,color={rgb, 255:red, 208; green, 2; blue, 27 }  ,opacity=1 ]  {$68$};
		\draw (166,96.4) node [anchor=north west][inner sep=0.75pt]  [font=\tiny,color={rgb, 255:red, 245; green, 166; blue, 35 }  ,opacity=1 ]  {$64$};
		\draw (211,28.4) node [anchor=north west][inner sep=0.75pt]  [font=\tiny,color={rgb, 255:red, 245; green, 166; blue, 35 }  ,opacity=1 ]  {$64$};
		\draw (282,42.4) node [anchor=north west][inner sep=0.75pt]  [font=\tiny,color={rgb, 255:red, 245; green, 166; blue, 35 }  ,opacity=1 ]  {$64$};
		\draw (339,25.4) node [anchor=north west][inner sep=0.75pt]  [font=\tiny,color={rgb, 255:red, 208; green, 2; blue, 27 }  ,opacity=1 ]  {$68$};
		\draw (185,183.4) node [anchor=north west][inner sep=0.75pt]  [font=\tiny,color={rgb, 255:red, 126; green, 211; blue, 33 }  ,opacity=1 ]  {$49.7$};
		\draw (244,75.4) node [anchor=north west][inner sep=0.75pt]  [font=\tiny,color={rgb, 255:red, 245; green, 166; blue, 35 }  ,opacity=1 ]  {$64$};
		\draw (171,125.4) node [anchor=north west][inner sep=0.75pt]  [font=\tiny,color={rgb, 255:red, 245; green, 166; blue, 35 }  ,opacity=1 ]  {$64$};
		\draw (266,121.4) node [anchor=north west][inner sep=0.75pt]  [font=\tiny,color={rgb, 255:red, 245; green, 166; blue, 35 }  ,opacity=1 ]  {$64$};
		\draw (300,93.4) node [anchor=north west][inner sep=0.75pt]  [font=\tiny,color={rgb, 255:red, 245; green, 166; blue, 35 }  ,opacity=1 ]  {$64$};
		\draw (346,113.4) node [anchor=north west][inner sep=0.75pt]  [font=\tiny,color={rgb, 255:red, 248; green, 231; blue, 28 }  ,opacity=1 ]  {$62.4$};
		\draw (207,144.4) node [anchor=north west][inner sep=0.75pt]  [font=\tiny,color={rgb, 255:red, 65; green, 117; blue, 5 }  ,opacity=1 ]  {$48$};
		\draw (255,174.4) node [anchor=north west][inner sep=0.75pt]  [font=\tiny,color={rgb, 255:red, 65; green, 117; blue, 5 }  ,opacity=1 ]  {$48$};
		\draw (314,154.4) node [anchor=north west][inner sep=0.75pt]  [font=\tiny,color={rgb, 255:red, 245; green, 166; blue, 35 }  ,opacity=1 ]  {$64$};
		\draw (360,227.4) node [anchor=north west][inner sep=0.75pt]  [font=\tiny,color={rgb, 255:red, 245; green, 166; blue, 35 }  ,opacity=1 ]  {$64$};
		\draw (348,165.4) node [anchor=north west][inner sep=0.75pt]  [font=\tiny,color={rgb, 255:red, 245; green, 166; blue, 35 }  ,opacity=1 ]  {$64$};
		\draw (400,146.4) node [anchor=north west][inner sep=0.75pt]  [font=\tiny,color={rgb, 255:red, 245; green, 166; blue, 35 }  ,opacity=1 ]  {$64$};
		\draw (406,207.4) node [anchor=north west][inner sep=0.75pt]  [font=\tiny,color={rgb, 255:red, 245; green, 166; blue, 35 }  ,opacity=1 ]  {$64$};
		\draw (453,177.4) node [anchor=north west][inner sep=0.75pt]  [font=\tiny,color={rgb, 255:red, 245; green, 166; blue, 35 }  ,opacity=1 ]  {$64$};
		\draw (237,254.4) node [anchor=north west][inner sep=0.75pt]  [font=\tiny,color={rgb, 255:red, 129; green, 3; blue, 3 }  ,opacity=1 ]  {$72$};
		\draw (299,257.4) node [anchor=north west][inner sep=0.75pt]  [font=\tiny,color={rgb, 255:red, 129; green, 3; blue, 3 }  ,opacity=1 ]  {$72$};
		\draw (292,214.4) node [anchor=north west][inner sep=0.75pt]  [font=\tiny,color={rgb, 255:red, 129; green, 3; blue, 3 }  ,opacity=1 ]  {$72$};
		\draw (341,68) node [anchor=north west][inner sep=0.75pt]  [font=\scriptsize] [align=left] {\mbox{-} Flows in MW};

	\end{tikzpicture}

	\captionof{figure}{{Flows, equilibrium prices and capacity bottlenecks in the Italian power network. Different colors highlight groups of markets that have different prices at equilibrium. Wavy links denote saturated power lines connecting these groups and the arrows indicate the actual direction of the energy flow. }}
	\label{fig4}
	
	\medskip
	
		\captionof{table}{{Price groups at equilibrium in the Italian power network. }}
	\label{tab1}
	\quad \quad \scalebox{0.9}{
		\begin{tabular}{|c|c|}
			\hline
			\textbf{Eq. Prices $\frac{\text{\euro}}{\text{MWh}}$} & \textbf{Markets}                \\ \hline
			72     & 20,21,22                        \\ \hline
			68     & 1,6                             \\ \hline
			64     & 2,3,4,5,7,8,9,13,14,16,17,18,19 \\ \hline
			62.4   & 16                              \\ \hline
			49.7   & 11                              \\ \hline
			48     & 12,15                           \\ \hline
		\end{tabular}%
	}

	\bigskip
	
	Notice that a total of 6 price groups of markets arise, each characterized by a different price at equilibrium. By Corollary \ref{col1}, we know that the power lines connecting these different groups must be saturated and the energy only flows in a certain direction, this is exactly what we observe numerically. In Fig \ref{fig4} we show the groups with different colors and the indication of the corresponding equilibrium price next to each node while the wavy links denote saturated lines. The weights on links denote the flow going through that line and arrows give tits actual direction.
	
	From the equilibrium state, we observe that higher prices are found in those markets with few producers and that are also sufficiently far from the main distribution hubs or directly cut out by severe capacity bottlenecks.
	Interestingly, we observe that even by choosing the same price and cost functions, price groups do not need to be connected components of the corresponding graph: in other words, there might be markets with the same price at equilibrium but not directly connected by a power line (see node 1 and 6 in this example). This is an effect due to the homogeneity of the parameters chosen for this example and we do not expect it to happen with fully general price and cost functions.

\section{Conclusions} \label{sec4}

In this paper we have studied a model of a networked Cournot competition involving producers and a market maker competing on multiple markets connected by links with finite capacity. This model is suited to describe energy marketplaces where the links connecting the markets represent physical power lines. We proved a very general result concerning the optimal action of the market maker and the presence of saturated cuts in the power network. This result allowed us to shed light on the implications of capacity bottlenecks in the power network on the emergence of price differences between different markets. Moreover, under mild assumptions on the utilities, we have studied the existence and uniqueness of the Nash equilibria of the proposed game.

Ongoing research is focused on exploiting our result on saturated cuts to develop optimal network intervention/design policies. Possible problems involve finding the critical cut and how to optimally create new lines or allocate additional capacity among the lines of the power network in order to level price differences or maximize certain welfare functions.

\bibliography{bib}

\end{document}